\documentclass[fleqn,12pt,twoside]{article}
\usepackage{espcrc1}
\usepackage{bm}

\usepackage{graphicx}
\usepackage[figuresright]{rotating}

\newcommand{\AmS}{{\protect\the\textfont2
  A\kern-.1667em\lower.5ex\hbox{M}\kern-.125emS}}

\title{Numerical study of the hadron-quark mixed phase 
       }

\author{Tomoki Endo\address[KYOTO]{Department of Physics, Kyoto University,  
        Kyoto, 606-8502, Japan \\ 
        }%
        \thanks{endo@ruby.scphys.kyoto-u.ac.jp},
        Toshiki Maruyama\address[GENKEN]{Advanced Science Research Center, Japan
        Atomic Energy Research Institute, Tokai, Ibaraki, 319-1195, Japan \\
        },
        Satoshi Chiba\addressmark[GENKEN]
        and
        Toshitaka Tatsumi\addressmark[KYOTO]
       }
\begin{document}

\maketitle

\begin{abstract}
 The Coulomb screening effect and the finite-size effect such as
 surface tension are figured out in the hadron-quark deconfinement phase transition.  
 We study the mixed phase of the quark droplets immersed in hadron
 matter. We see that the droplet phase is
 mechanically unstable if the surface tension is strong enough. 
 Once the Coulomb potential is properly taken into account, we could
 effectively satisfy the condition for charge chemical equilibrium in the
 Maxwell construction. As a result, we suggest the Maxwell construction  revives the physical
 meaning effectively.
\end{abstract}

\section{Introduction}

 Nowadays, it has been believed that hadron matter becomes quark matter in high-density
 region, that is to say ``deconfinement phase transition''. We are here 
 interested in how 
 hadron matter changes to quark matter at zero temperature.

 In the first-order phase transition, such as the water-vapor phase transition, we know the
 the Maxwell construction (MC) works to derive the equation of state in
 thermodynamic equilibrium.
 When we naively apply MC to this deconfinement phase transition, we immediately
 find one of the Gibbs conditions (GC) is not fulfilled,
 which are basic conditions in thermodynamics;
\begin{equation}
 \mu_{\mathrm{B}}^{\mathrm{quark}} = \mu_{\mathrm{B}}^{\mathrm{hadron}} (\equiv \mu_{\mathrm{B}} ),
 \hspace{5pt} \mu_{\mathrm{e}}^{\mathrm{quark}} \neq
 \mu_{\mathrm{e}}^{\mathrm{hadron}} , \hspace{5pt} P^{\mathrm{quark}}=P^{\mathrm{hadron}} , \hspace{5pt} T^{\mathrm{quark}} = T^{\mathrm{hadron}} 
\end{equation}
where $\mu_{\mathrm{B}}^i$ and $\mu_{\mathrm{e}}^i$ are baryon-number
and charge chemical potentials, respectively.

 Glendenning \cite{gle} claimed, by the bulk calculation, that MC is not appropriate in quark-hadron
 deconfinement phase transition because of more than one chemical
 potential, and that there is assumed local charge
 neutrality instead of total charge neutrality; both hadron matter and quark matter
should be charged. Consequently there appears no constant-pressure region like in
MC and the structured mixed-phase develops in a wide density
region.

The bulk calculation assumes bulk quark matter and hadron matter without
  any
 finite-size effect like the surface tension and the Coulomb interaction.
Therefore matter is no more uniform but takes geometrical
structures in the mixed phase. These geometrical structures are called ``droplet'', ``rod'', ``slab'', ``tube'' and ``bubble.''

On the other hand, Heiselberg et al \cite{pet} pointed out that 
the mixed phase should appear in a narrow density region and it is energetically disfavored for a large
value of the surface tension. In the previous work \cite{vos} it has been 
pointed out that we could satisfy the condition of charge chemical equilibrium even in MC case  
by properly introducing the Coulomb potential, 

\begin{equation}
\hfill \mu_{\mathrm{e}}^{\mathrm{quark}}\ = \mu_{\mathrm{e}}^{\mathrm{hadron}} ( \equiv \mu_{\mathrm{e}} ) \hfill
\end{equation}
where $\mu_{\mathrm{e}}^{\mathrm{quark}} - V_{\mathrm{Coul}}^{\mathrm{quark}} \neq
 \mu_{\mathrm{e}}^{\mathrm{hadron}} - V_{\mathrm{Coul}}^{\mathrm{hadron}}$
which corresponds to Eq.~(1) in  a gauge invariant fashion.

 We can  see  that the Coulomb screening
effect plays an important role and that the
geometrical structures become mechanically unstable in the hadron-quark deconfinement phase  transition.
However, there is used the linearized approximation for the Poisson
equation to figure out the Coulomb screening effect analytically.  

In this study, we numerically solve the Poisson equation without any
approximation. As mentioned above, since the Coulomb potential and
its Coulomb screening effect should be so important to change the description of
the quark-hadron mixed phase, we must carefully treat the Coulomb interaction.

\section{Formalism}

We use the density functional theory (DFT) to discuss the quark-hadron
mixed phase \cite{vos}.
Consider 
quark droplets with 
a
radius $R$, embedded in nuclear matter.
We divide the whole space into equivalent Wigner-Seitz cells with 
a
radius
$R_{\mathrm{W}}$, and impose the total
charge neutrality and chemical
equilibrium in the quark and hadron phases and at the quark-hadron boundary. 
The quark phase consists of $u$, $d$, $s$ quarks and
electrons in chemical equilibrium. To take into account the confinement
we  incorporate
the bag model since we poorly understand the physical picture of the hadron-quark
interface. We use the bag constant $B=120 \hspace{3pt} \mathrm{MeV/fm^3}$,
which is
equal to that in Ref. \cite{pet}. As for the hadron phase it consists of nucleons and electrons in
 chemical equilibrium. We use
an effective potential parametrized to reproduce the saturation property of nuclear
matter.

At the interface of the quark and hadron phases, we introduce a sharp boundary on the
basis of the bag model with the surface
tension $\sigma$. Its value should depend on
the bag constant and other values such as $s$ quark mass, while we treat
it as a free parameter in this study.

Differentiating the thermodynamic potential $\Omega$ with respect to each constituent
density $\rho_i \hspace{3pt} (i=u,d,s,n,p,e)$ or the Coulomb potential, $V_{\mathrm{Coul}}$, we obtain the equations of motion
(EOM) for $\rho_i$  and for $V_{\mathrm{Coul}}$. We numerically solve these EOM
with the conditions of chemical equilibrium in each phase and at hadron-quark boundary;
especially, we fully solve the
Poisson equation without any approximation. Note that the  Poisson equation
becomes highly non-linear because particle charge densities are complicated functions
of the Coulomb potential.

\newpage

\begin{figure}[htb]
\begin{minipage}[t]{80mm}
  \includegraphics[width=70mm]{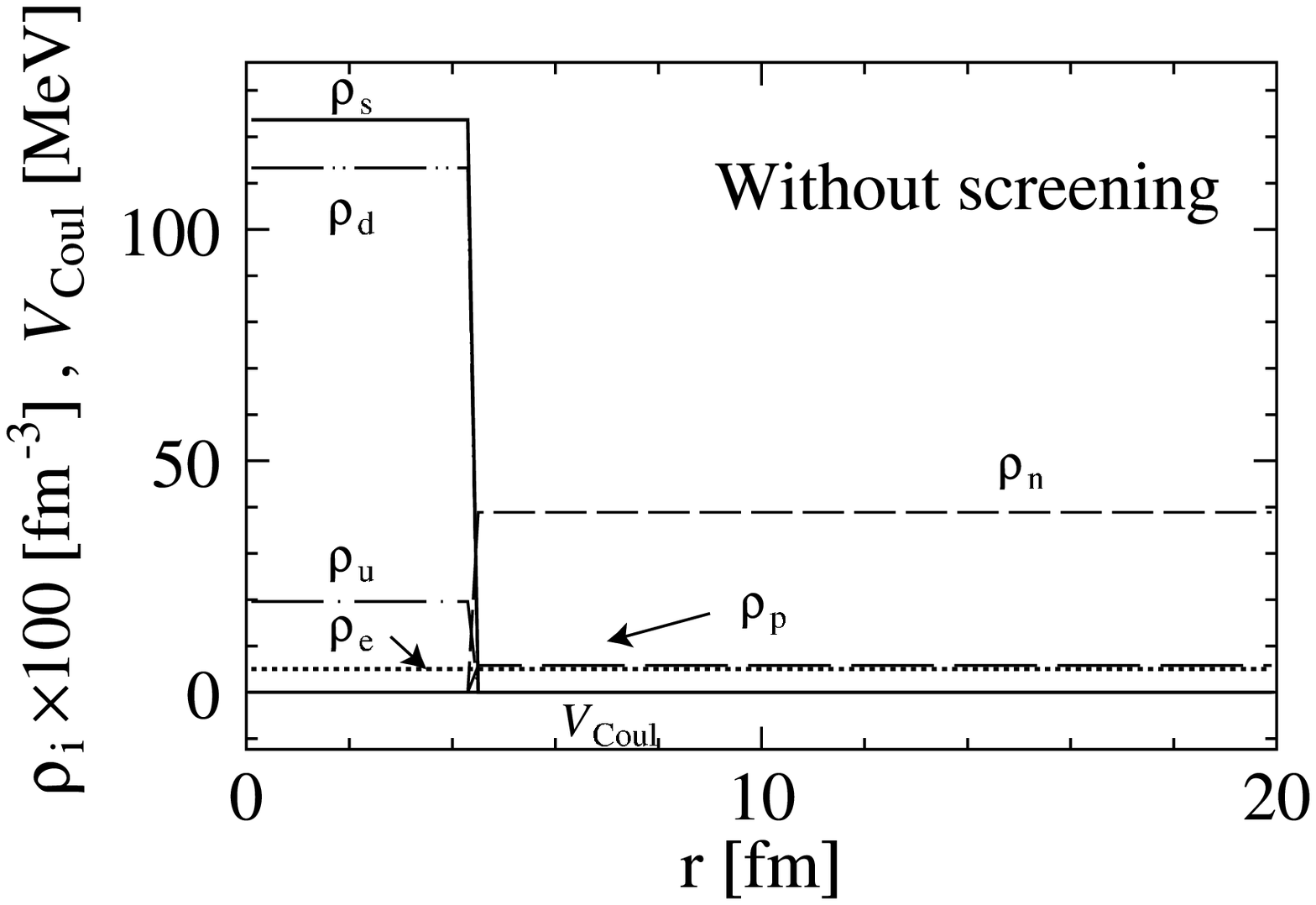}
\end{minipage}
\hspace{8pt}
\begin{minipage}[t]{80mm}
  \includegraphics[width=70mm]{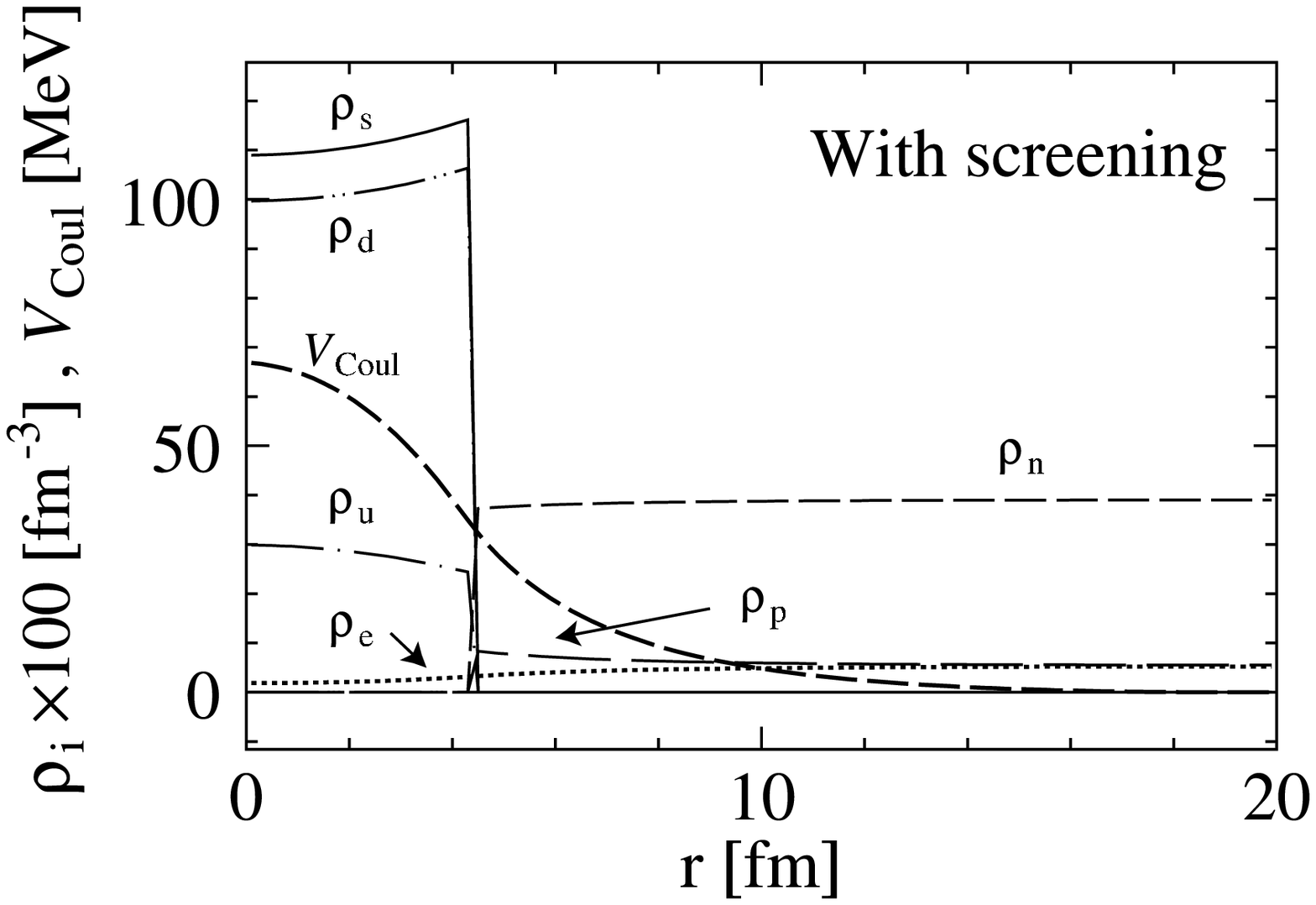}
\end{minipage}
\caption{Density profiles $\rho_i$ without the Coulomb screening effect (left panel) and with
 the Coulomb screening effect (right panel). This is the case of
 $\mu_{\mathrm{B}}=1232$ MeV and the volume fraction
 $f=\left(R/R_{\mathrm{W}}\right)^3 = 0.01$. Each density uniformly
 spreads in each phase and $V_{\mathrm{Coul}}$ is constant ( =0 ) in the
 left panel, while there is re-arrangement of the charge densities and
 $V_{\mathrm{Coul}}$ is spatially dependent in the right panel.}
\end{figure}

\section{Numerical results}

 Figure 1 shows one example of the density profiles without and with the
Coulomb screening effect. 
In the case without the Coulomb screening,
each density is uniform in each phase (left panel), while it is
not uniform 
when 
the Coulomb screening effect 
is taken into account
(right panel). Most remarkable
difference between these two figures 
lies 
near the hadron-quark
interface and the center of the droplet. Negatively charged particle
densities are suppressed as they are apart from the boundary; $d$, $s$ and $e$
in the quark phase and $e$ in the hadron phase. On the contrary,
positively charged ones are enhanced; $u$ in the quark phase and $p$ in the
hadron phase. As a result, each phase tends to be charge neutral by
itself, which should resemble the picture given by MC.

\begin{figure}[htb]
\begin{minipage}[t]{0.48\textwidth}
  \includegraphics[width=67mm]{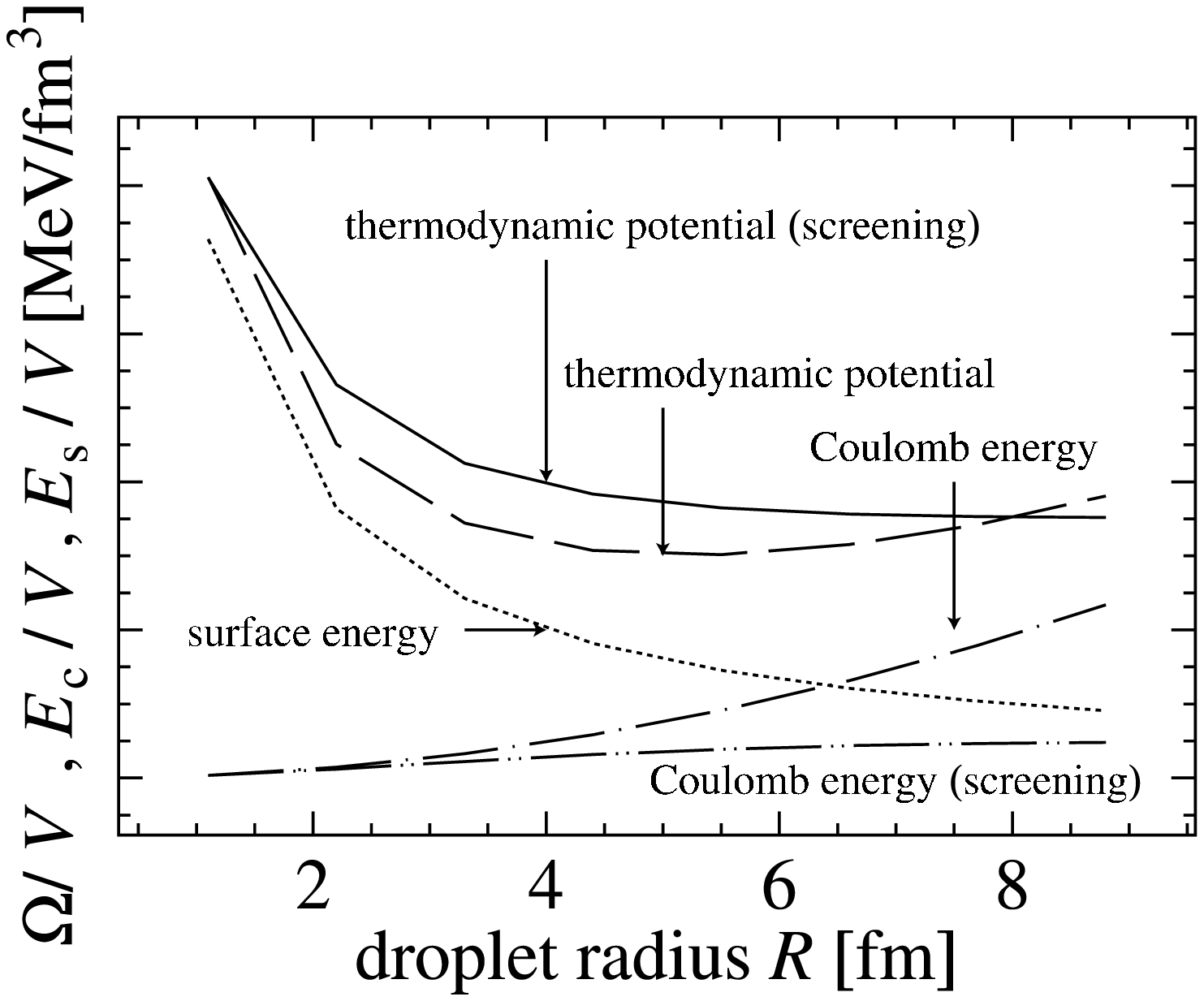}
\caption{The thermodynamic potential and energy contributions as functions
 of the droplet radius $R$. This is the case of $\mu_{\mathrm{B}}=1232$ MeV,
 $f=0.01$ and $\sigma=60 \hspace{3pt} \mathrm{MeV/fm^2}$. We can see the Coulomb
 energy is suppressed by the Coulomb screening
 effect for large $R$ and the minimum of the thermodynamic potential disappears, which shows
 the mechanical instability.}
\end{minipage}
\hspace{8pt}
%
\begin{minipage}[t]{0.48\textwidth}
  \includegraphics[width=67mm]{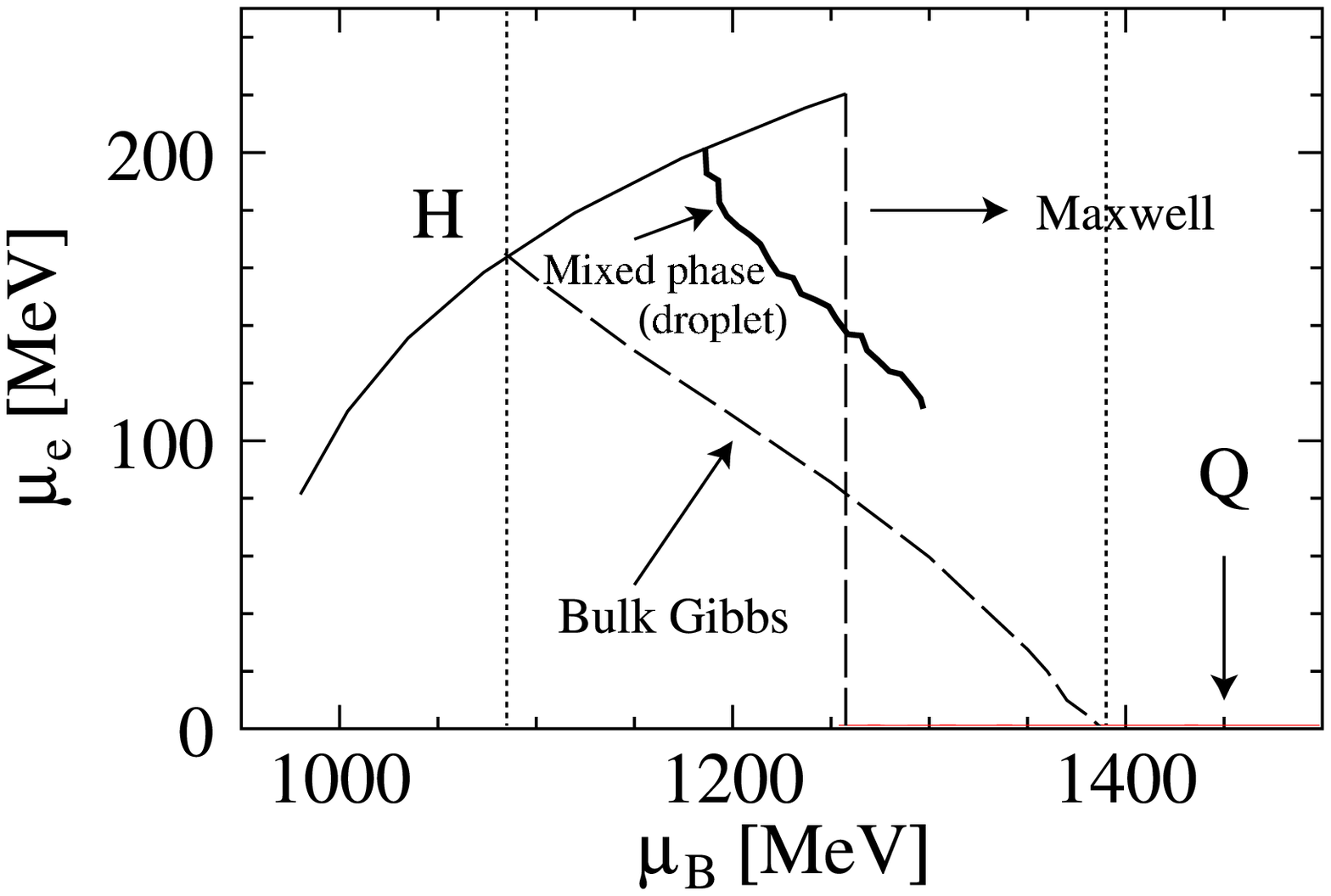}
\caption{Phase diagram in the  $\mu_{\mathrm{B}}-\mu_{\mathrm{e}}$ plane
 with $\sigma=40 \hspace{3pt} \mathrm{MeV/fm^2}$. In this case,
  the $\mu_{\mathrm{B}}-\mu_{\mathrm{e}}$ curve for the droplet phase can
 appear far from ``Bulk Gibbs''. Note that
 the $\mu_{\mathrm{B}}-\mu_{\mathrm{e}}$ curve for uniform quark matter denoted by ``Q'' is lying fairly close to horizontal axis because
 $\mu_e \sim 0$ in uniform quark matter.
 The $\mu_{\mathrm{B}}-\mu_{\mathrm{e}}$ curve for uniform hadron matter
 is denoted by ``H''.}
\label{fig:toosmall}
\end{minipage}
\vspace*{-5mm}
\end{figure}

As has been suggested in Ref. \cite{vos}, the Coulomb screening effect may
induce the mechanical instability. In Fig.~2 we demonstrate an example 
of the case with strong surface tension
in our framework.
 We can see that the Coulomb screening effect strongly suppresses the Coulomb
 energy for large $R$ and thereby minimum of the thermodynamic potential disappears,
which means droplet is mechanically unstable.

We show the phase diagram for the ``droplet'' phase in Fig.~3, 
where the droplet can exist as an energetically 
%
favored
 and mechanically stable state. We also depict two curves denoted by
 ``Maxwell'' and ``Bulk Gibbs'' for comparison, which imply the one given
 by MC and the one constructed by GC under the ansatz of bulk quark
 matter and hadron matter, respectively. We can see the 
 $\mu_{\mathrm{B}}-\mu_{\mathrm{e}}$ curve for the droplet phase appears far from ``Bulk Gibbs''.
 This feature resembles the case of kaon condensation \cite{mar}. 
While the droplet phase is just one of the structured mixed-phases, 
it could be said that even when there appear other geometrical
structures, its
$\mu_{\mathrm{B}}-\mu_{\mathrm{e}}$ curve exhibits the similar behavior
 ``Maxwell''.  Hence, the Maxwell construction
may revive the physical meaning effectively.

\section{Summary}

We have seen the Coulomb screening effect and the finite-size effect in the hadron-quark deconfinement phase
transition. We have considered here the droplet phase as one of the
structured mixed-phases, and found that droplet is mechanically unstable when the surface tension is strong enough.
 We have found that the Coulomb screening effect as well as the finite-size effect work against the structured mixed-phase and thereby
restrict the mixed phase in a narrow density region.
 It could be suggested, from our work, that there cannot exist wide
 mixed phase region but only narrow region in neutron star core.
However, we have to study other geometrical
structures of ``rod'', ``slab'', ``tube'' and ``bubble'' \cite{end} than ``droplet''. After this task we can complete the phase diagram in the $\mu_{\mathrm{B}}-\mu_{\mathrm{e}}$ plane.

\end{document}